\documentclass{article}
\usepackage{spconf,amsmath,graphicx, hyperref,xcolor,acronym, comment}
\usepackage{colortbl}
\usepackage{hyperref}
\hypersetup{
    colorlinks=true,
    linkcolor=blue,
    filecolor=magenta,      
    urlcolor=cyan,
}
\usepackage{spconf,amsmath}

\title{ICASSP 2021 Deep Noise Suppression Challenge}
\name{Chandan K. A. Reddy, Harishchandra Dubey, Vishak Gopal, Ross Cutler, Sebastian Braun,}
\secondlinename{Hannes Gamper, Robert Aichner, Sriram Srinivasan}
\address{Microsoft Corporation, Redmond, USA}

\begin{document}
\ninept

\maketitle

\begin{abstract}
The Deep Noise Suppression (DNS) challenge is designed to foster innovation in the area of noise suppression to achieve superior perceptual speech quality. We recently organized a DNS challenge special session at INTERSPEECH 2020. We open-sourced training and test datasets for researchers to train their noise suppression models. We also open-sourced a subjective evaluation framework and used the tool to evaluate and pick the final winners. Many researchers from academia and industry made significant contributions to push the field forward. We also learned that as a research community, we still have a long way to go in achieving excellent speech quality in challenging noisy real-time conditions. In this challenge, we are expanding both our training and test datasets. Clean speech in the training set has increased by 200\% with the addition of singing voice, emotion data, and non-English languages. The test set has increased by 100\% with the addition of singing, emotional, non-English (tonal and non-tonal) languages, and, personalized DNS test clips. There are two tracks with a focus on (i) real-time denoising, and (ii) real-time personalized DNS. 
\end{abstract}

\vspace{-2mm}
\section{Introduction}
In recent times, remote work has become the "new normal" as the number of people working remotely has exponentially increased due to the pandemic. There has been a surge in the demand for reliable collaboration and real-time communication tools. Audio calls with very good to excellent speech quality are needed during these times as we try to stay connected and collaborate with people every day. We are easily exposed to a variety of background noises such as a dog barking, a baby crying, kitchen noises, etc. Background noise significantly degrades the quality and intelligibility of the perceived speech leading to fatigue. Background noise poses a challenge in other applications such as hearing aids and smart devices. 

Real-time Speech Enhancement (SE) for perceptual quality is a decades old classical problem and researchers have proposed numerous solutions~\cite{malah, 8031044}. In recent years, learning-based approaches have shown promising results~\cite{8281993, choi2020phase, koyama2020exploring}. The Deep Noise Suppression (DNS) Challenge organized at INTERSPEECH 2020 showed promising results, while also indicating that we are still about 1.4 Differential Mean Opinion Score (DMOS) from the ideal Mean Opinion Score (MOS) of 5 when tested on the DNS Challenge test set~\cite{valin2020perceptually, isik2020poconet}. The DNS Challenge is the first contest that we are aware of that used subjective evaluation to benchmark SE methods using a realistic noisy test set \cite{reddy2020interspeech}. We open sourced clean speech and noise corpus with configurable scripts to generate noisy-clean speech pairs suitable to train a supervised noise suppression model. There were two tracks, real-time and non-real-time based on the computational complexity of the inference. We received an overwhelming response to the challenge with participation from a diverse group of researchers, developers, students, and hobbyists from both academia and industry. We also received positive responses from the participants as many found the open sourced datasets quite useful, and both the dataset and test framework have been cloned at a fast rate since the challenge.    

The ICASSP 2021 Deep Noise Suppression (DNS) Challenge\footnote{\url{https://github.com/microsoft/DNS-Challenge}} is intended to stimulate research in the area of real-time noise suppression. For ease of reference, we will call the ICASSP 2021 challenge as DNS Challenge 2 and the Interspeech 2020 challenge as DNS Challenge 1. The DNS Challenge 2 will have a real-time denoising track similar to the one in DNS Challenge 1. In addition, we will have a personalized DNS track focused on using speaker information to achieve better perceptual quality. In addition to the datasets we open sourced for DNS Challenge 1, we increased clean speech in training set by 50\% resulting in over 760 hours which includes singing voice, emotion data, and non-english languages (Chinese). Noise data in training set remains the same as DNS Challenge 1. We provide over 118,000 room impulse responses (RIR), which includes real and synthetic RIRs from public datasets. We provide acoustic parameters: Reverberation time (T60) and Clarity (C50) for read clean speech and RIR sample. In DNS Challenge 2, we increased the testset by 100\% by adding emotion data, singing voice, non-English languages in Track 1 and real and synthetic clips for personalized DNS in Track 2. For DNS Challenge 1, we open sourced a subjective evaluation framework based on ITU-T P.808 \cite{naderi2020open}. 
The final evaluation of the participating models were done based on subjective evaluation using the P.808 subjective testing framework. We describe the results of the challenge at the end.
\vspace{-2mm}
\section{Challenge Tracks}
\label{challenge_tracks}
\vspace{-1mm}
The challenge had the following two tracks:

\begin{enumerate}
    \item Track 1: Real-Time Denoising track requirements
    \begin{itemize}
        \item The noise suppressor must take less than the stride time $T_s$ (in ms) to process a frame of size $T$ (in ms) on an Intel Core i5 quad-core machine clocked at 2.4 GHz or equivalent processors. For example, $T_s = T/2$ for 50\% overlap between frames. The total algorithmic latency allowed including the frame size $T$, stride time $T_s$, and any look ahead must be $\leq$ 40ms. For example, for a real-time system that receives 20ms audio chunks, if you use a frame length of 20ms with a stride of 10ms resulting in an algorithmic latency of 30ms, then you satisfy the latency requirements. If you use a frame size of 32ms with a stride of 16ms resulting in an algorithmic latency of 48ms, then your method does not satisfy the latency requirements as the total algorithmic latency exceeds 40ms. If your frame size plus stride $T_1=T+T_s$ is less than 40ms, then you can use up to $(40-T_1)$ms future information.   
    \end{itemize}
    \item Track 2:  Personalized Deep Noise Suppression (pDNS) track requirements
    \begin{itemize}
        \item Satisfy Track 1 requirements.
        \item You will have access to 2 minutes speech of a particular speaker to extract and adapt speaker related information that might be useful to improve the quality of the noise suppressor. The enhancement must be done on the noisy speech test segment of the same speaker. 
        \item The enhanced speech using speaker information must be of better quality than enhanced speech without using the speaker information.
    \end{itemize}
\end{enumerate}
\vspace{-2mm}
\section{Training Datasets}
The goal of releasing the clean speech and noise datasets is to provide researchers with an extensive and representative dataset to train their SE models. We initially released MSSNSD \cite{reddy2019scalable} with a focus on extensibility, but the dataset lacked the diversity in speakers and noise types. We published a significantly larger and more diverse data set with configurable scripts for DNS Challenge 1 \cite{reddy2020interspeech}. Many researchers found this dataset useful to train their noise suppression models and achieved good results. However, the training and the test datasets did not include clean speech with emotions such as crying, yelling, laughter or singing. Also, the dataset only includes the English language. For DNS Challenge 2, we are adding speech clips with other emotions and included about 10 non-English languages. Clean speech in training set is total 760.53 hours: read speech (562.72 hours), singing voice (8.80 hours), emotion data (3.6hours), Chinese mandarin data (185.41 hours). We have grown clean speech to 760.53 hours as compared to 562.72 hours in DNS Challenge 1. The details about the clean and noisy dataset are described in the following sections.

\subsection{Clean Speech}
\label{ssec:cleanspeech}
Clean speech consists of three subsets: (i) Read speech recorded in clean conditions; (ii) Singing clean speech; (iii) Emotional clean speech; and (iv) Non-english clean speech. The first subset is derived from the public audiobooks dataset called Librivox\footnote{https://librivox.org/}. It is available under the permissive creative commons 4.0 license \cite{7178964}. It has recordings of volunteers reading over 10,000 public domain audiobooks in various languages, the majority of which are in English. In total, there are 11,350 speakers. Many of these recordings are of excellent speech quality, meaning that the speech was recorded using good quality microphones in a silent and less reverberant environments. But there are many recordings that are of poor speech quality as well with speech distortion, background noise, and reverberation. Hence, it is important to clean the data set based on speech quality. We used the online subjective test framework ITU-T P.808 \cite{naderi2020open} to sort the book chapters by subjective quality. The audio chapters in Librivox are of variable length ranging from few seconds to several minutes. We randomly sampled 10 audio segments from each book chapter, each of 10 seconds in duration. For each clip, we had 2 ratings, and the MOS across all clips was used as the book chapter MOS. Figure 1 shows the results, which show the quality spanned from very poor to excellent quality. In total, it is 562 hours of clean speech, which was part of DNS Challenge 1.

The second subset consists of high-quality audio recordings of singing voice recorded in noise-free conditions by professional singers. This subset is derived from~\textit{VocalSet} corpus~\cite{wilkins2018vocalset} with Creative Commons Attribution 4.0 International License (CC BY 4.0). license. It has 10.1 hours of clean singing voice recorded by 20 professional singers: 9 males, and 11 females. This data was recorded on a range of vowels, a diverse set of voices on several standard and extended vocal techniques, and sung in contexts of scales, arpeggios, long tones, and excerpts. We downsampled the mono .WAV files from 44.1kHz to 16kHz and added it to clean speech used by the training data synthesizer. 

The third subset consists of emotion speech recorded in noise-free conditions. This is derived from Crowd-sourced Emotional Mutimodal Actors Dataset (CREMA-D)~\cite{cao2014crema} made available under the Open Database License. It consists of 7,442 audio clips from 91 actors: 48 male, and 43 female accounting to total 3.5 hours of audio. The age of the actors was in the range of 20 to 74 years with diverse ethnic backgrounds including African America, Asian, Caucasian, Hispanic, and Unspecified. Actors read from a pool of 12 sentences for generating this emotional speech dataset. It accounts for six emotions: Anger, Disgust, Fear, Happy, Neutral, and Sad at four intensity levels: Low, Medium, High, Unspecified. The recorded audio clips were annotated by multiple human raters in three modalities: audio, visual, and audio-visual. Categorical emotion labels and real-value emotion level values of perceived emotion were collected using crowd-sourcing from 2,443 raters. This data was provided as 16 kHz .WAV files so we added it to our clean speech as it is. 

The fourth subset has clean speech from non-English languages. It consists of both tonal and non-tonal languages including Chinese (Mandarin), German and Spanish. Mandarin data consists of OpenSLR18~\footnote{http://www.openslr.org/18/} THCHS-30~\cite{THCHS30_2015} and OpenSLR33~\footnote{http://www.openslr.org/33/} AISHELL~\cite{aishell_2017} datasets, both with Apache 2.0 license. THCHS30 was published by Center for Speech and Language Technology (CSLT) at Tsinghua University for speech recognition. It consists of 30+ hours of clean speech recorded at 16-bit 16 kHz in noise-free conditions. Native speakers of standard Mandarin read text prompts chosen from a list of 1000 sentences. We added the entire THCHS-30 data in our clean speech for the training set. It consisted of 40 speakers: 9 male, 31 female in the age range of 19-55 years. It has total 13,389 clean speech audio files~\cite{THCHS30_2015}. The AISHELL dataset was created by Beijing Shell Shell Technology Co. Ltd. It has clean speech recorded by 400 native speakers ( 47\% male and 53\% female) of Mandarin with different accents. The audio was recorded in noise-free conditions using high fidelity microphones. It is provided as 16-bit 16kHz .wav files. It is one of the largest open-source Mandarin speech datasets. We added the entire AISHELL corpus with 141,600 utterances spanning 170+ hours of clean Mandarin speech to our training set.

Spanish data is 46 hours of clean speech derived from OpenSLR39, OpenSLR61, OpenSLR71, OpenSLR73, OpenSLR74 and OpenSLR75 where re-sampled all .WAV files to 16 kHz. German data is derived from four corpora namely (i) The Spoken Wikipedia Corpora~\cite{swc}, (ii) Telecooperation German Corpus for Kinect~\cite{kinect}, (iii) M-AILABS data~\cite{mailabs}, (iv) zamia-speech forschergeist corpora. Complete German data constitute 636 hours. Italian (128 hours), French (190 hours), Russian (47 hours) are taken from M-AILABS data~\cite{mailabs}. M-AILABS Speech Dataset is a  publicly available multi-lingual corpora for training speech recognition and speech synthesis systems.
\begin{figure}[!tb]
\includegraphics[width=0.8\columnwidth]{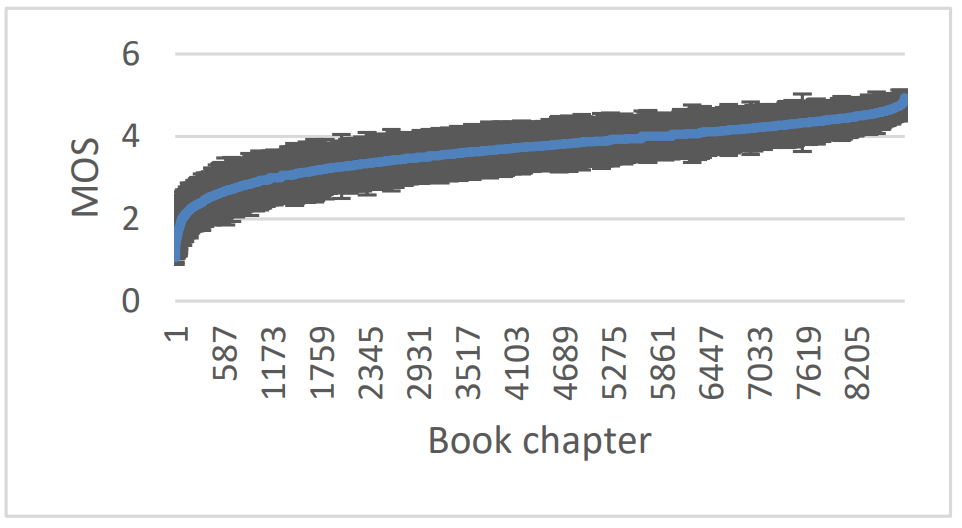}
\caption{Sorted near-end single-talk clip quality (P.808) with 95\% confidence intervals.}
\label{fig:nearend}
\end{figure}
\vspace{-1mm}
\subsection{Noise}
\label{ssec:noise}
 The noise clips were selected from Audioset \footnote{https://research.google.com/audioset/} \cite{7952261} and Freesound \footnote{https://freesound.org/}. Audioset is a collection of about 2 million human labeled 10s sound clips drawn from YouTube videos and belong to about 600 audio events. Like the Librivox data, certain audio event classes are over-represented. For example, there are over a million clips with audio classes music and speech and less than 200 clips for classes such as toothbrush, creak, etc. Approximately 42\% of the clips have a single class, but the rest may have 2 to 15 labels. Hence, we developed a sampling approach to balance the dataset in such a way that each class has at least 500 clips. We also used a speech activity detector to remove the clips with any kind of speech activity, to strictly separate speech and noise data. The resulting dataset has about 150 audio classes and 60,000 clips. We also augmented an additional 10,000 noise clips downloaded from Freesound and DEMAND databases \cite{demand}. The chosen noise types are more relevant to VOIP applications. In total, there is 181 hours of noise data.
\vspace{-1mm}
\subsection{Room Impulse Responses}
We provide 3076 real and approximately 115,000 synthetic rooms impulse responses (RIRs) where we can choose either one or both types of RIRs for convolving with clean speech. Noise is then added to reverberant clean speech while DNS models are expected to take noisy reverberant speech and produce clean reverberant speech. Challenge participants can do both de-reverb and denoising with their models if they prefer. These RIRs are chosen from openSLR26~\cite{ko2017study}~\footnote{http://www.openslr.org/26/} and openSLR28~\cite{ko2017study}~\footnote{http://www.openslr.org/28/} datasets, both released with Apache 2.0 License.
\vspace{-1mm}
\subsection{Acoustic parameters}
We provide two acoustic parameters: (i) Reverberation time, T60~\cite{antsalo2001estimation} and (ii) Clarity, C50~\cite{gamper2020blind} for all audio clips in clean speech of training set. We provide T60, C50 and isReal Boolean flag for all RIRs where isReal is 1 for real RIRs and 0 for synthetic ones. The two parameters are correlated. A RIR with low C50 can be described as highly reverberant and vice versa~\cite{antsalo2001estimation,gamper2020blind}. These parameters are supposed to provide flexibility to researchers for choosing a sub-set of provided data for controlled studies. 
\vspace{-2mm}
\section{Test set}
In DNS Challenge 1, the test set consisted of 300 real recordings and 300 synthesized noisy speech clips. The real clips were recorded internally at Microsoft and also using crowdsourcing tools. Some of the clips were taken from Audioset. The synthetic clips were divided into reverberant and less reverberant clips. These utterances were predominantly in English. All the clips are sampled at 16 kHz with an average clip length of 12 secs. The development phase test set is in the \textbf{"ICASSP\_dev\_test\_set"} directory in the DNS Challenge repository. For this challenge, the primary focus is to make the test set as realistic and diverse as possible. 
\vspace{-1mm}
\subsection{Track 1}
\label{ssec:track1}
Similar to DNS Challenge 1, the test set for DNS Challenge 2 is divided into real recordings and synthetic categories. However, the synthetic clips are mainly composed of the scenarios that we were not able to collect in realistic conditions. The track 1 test clips can be found in the \textbf{track\_1} sub-directory of \textbf{ICASSP\_dev\_test\_set}.
\vspace{-1mm}
\subsubsection{Real recordings}
\label{sssec:realrec}
The real recordings consist of non-English and English segments. The English segment will have 300 clips that are from the blind test set from DNS Challenge 1. These clips were collected using the crowdsourcing platform and internally at Microsoft using a variety of devices, acoustic and noisy conditions. The non-English segment comprises of 100 clips in the following languages: Non-tonal: Portuguese, Russian, Spanish, and Tonal: Mandarin, Cantonese, Punjabi, and Vietnamese. In total, there are 400 real test clips.  
\vspace{-1mm}
\subsubsection{Synthetic test set}
\label{sssec:synth}
The synthetic test set consists of 200  noisy clips obtained by mixing clean speech (non-English, emotional speech, and singing) with noise. The test set noises is taken from Audioset and Freesound~\cite{reddy2020interspeech}. The 100 non-English test clips include German, French and Italian languages from Librivox audio books.  Emotion clean speech consists of laughter, yelling, and crying chosen from Freesound and mixed with test set noise to generate 50 noisy clips. Similarly, clean singing voice from Freesound was used to generate 50 noisy clips for the singing test set. 
%
\subsubsection{Blind test set}
\label{sssec:blind}
The blind test set for track 1 contains 700 noisy speech clips out of which 650 are real recordings and 50 synthetic noisy singing clips. It contains the following categories: (i) emotional (102 clips), (ii) English (276 clips), (iii) non-English including tonal (272 clips), (iv) tonal languages (112 clips) and (v) singing (50 clips). The real recordings were collected using the crowdsourcing platform and internally at Microsoft. This is the most diverse publicly available test set for a noise suppression task.
\vspace{-2mm}
\subsection{Track 2}
\label{ssec:track2}
For the pDNS track, we provide 2 minutes of clean adaptation data for each primary speaker with the goal to suppress neighboring speakers and background noise. pDNS models are expected to leverage speaker-aware training and speaker-adapted inference. There are two motivations to provide clean speech for the primary speaker: (1) speaker models are sensitive to false-alarms in speech activity detection (SAD)~\cite{hansen2015speaker}; clean speech can be used for obtaining accurate SAD labels. (2) speaker adaptation is expected to work well using multi-conditioned data; clean speech can be used for generating reverberant and noisy data for speaker adaptation. 

\subsubsection{Real Recordings}
The development test set contains 100 real recordings from 20 primary speakers collected using crowdsourcing. Each primary speaker has noisy test clips for three scenarios: (i) primary speaker in presence of neighboring speaker; (ii) primary speaker in presence of background noise; and (iii) primary speaker in presence of both background noise and neighboring speaker. 

\subsubsection{Synthetic test clips}
The synthetic clips include 500 noisy clips from 100 primary speakers. Each primary speaker has 2 minutes of clean adaptation data. All clips have varying levels of neighboring speakers and noise. test set noise from Track 1 was mixed with primary speech extracted from VCTK corpus~\cite{yamagishi2019cstr}. We used VoxCeleb2~\cite{chung2018voxceleb2} corpus for neighboring speakers.

\subsubsection{Blind test set}
\label{sssec:blind}
The blind test set for track 2 contains 500 noisy speech real recordings from 80 unique speakers. All the real recordings were collected using the crowdsourcing platform. The noise source in the majority of these clips is a secondary speaker. We provided 2 mins clean speech utterances for each of the primary speakers that could be used to adapt the noise suppressor. All the utterances were in English.

%
%
\vspace{-2mm}
\section{Challenge Results}
\vspace{-1mm}
\subsection{Evaluation Methodology}
Most DNS evaluations use objective measures such as and PESQ \cite{p862}, SDR, and POLQA \cite{polqa}. However, these metrics are shown to not correlate well with subjective speech quality in the presence of background noise \cite{Avila2019}. Subjective evaluation is the gold standard for this task. Hence, the final evaluation was done on the blind test set using the crowdsourced subjective speech quality metric based on ITU P.808 \cite{naderi2020open} to determine the DNS quality. For track 2, we appended each processed clip with a 5 secs utterance of the primary speaker at the beginning and 1 sec silence. We modified P.808 slightly to instruct the raters to focus on the quality of the voice of the primary speaker in the remainder of the processed segment. We conducted a reproducibility test with this change to P.808 and found that the average Spearman Rank Correlation Coefficient (SRCC) between the 5 runs was 0.98. Hence, we concluded that the change is valid. We used 5 raters per clip. This resulted in a 95\% confidence interval (CI) of 0.03. We also provided the baseline noise suppressor \cite{braun2020data} for the participants to benchmark their methods.  
\vspace{-2mm}
\begin{table}[h!]
  \begin{center}
    \caption{Track 1 P.808 Results.}
    \label{tab:table1}
    \includegraphics[width=0.9\columnwidth]{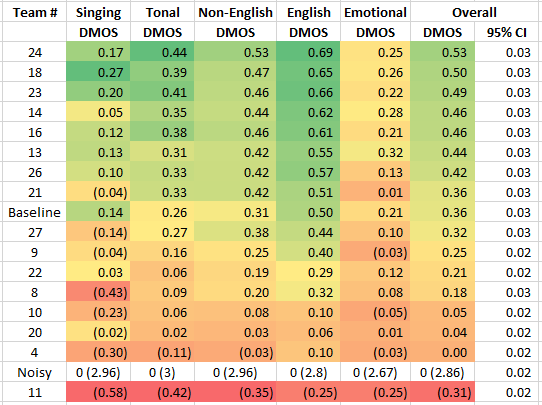}
    \end{center}
\end{table}
\subsection{Evaluation Results}
\subsubsection{Track 1}
A total of 16 teams from academia and industry participated in track 1. Table \ref{tab:table1} shows the categorized P.808 results for track 1. The submissions are stack ranked based on the overall Differential MOS (DMOS). DMOS is the difference in MOS between the processed set and the original noisy. We can observe from the results that most of methods struggled to do well on singing and emotional clips. The noise suppressors tend to suppress the certain emotional and singing segments. Overall, the results show that performance of noise suppressors are not great in singing and emotional categories. The results also shows that the training data must balanced between English, non-English and tonal languages for the model to generalize well. It is important to include singing and other emotions as the ground truth to achieve better quality in these categories. Only half of the submissions did better than the baseline. The best model is only 0.53 DMOS better than the noisy, which is of absolute MOS 3.38. This shows that we are far away from achieving a noise suppressor that works robustly in almost all conditions.  

\subsubsection{Track 2}
This track is the first of its kind and there is not much work done in this field. There were only 2 teams who participated in track 2. Each team submitted one set of processed clips with using speaker information for model adaptation and the other set without explicitly using speaker information. The results are shown in table  \ref{tab:table2}. The best model gave only 0.14 DMOS. This shows that the problem of using speaker information to adapt the model is still in the infancy stage. 
\vspace{-2mm}
\begin{table}[t!]
  \begin{center}
    \caption{P.808 Results for Track 2}
    \label{tab:table2}
    \includegraphics[width=0.6\columnwidth]{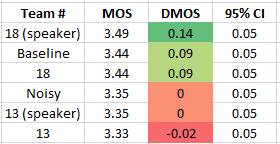}
    \end{center}
\end{table}
\vspace{0mm}
\section{Summary~\& Conclusions}
\vspace{-1mm}
The ICASSP 2021 DNS Challenge was designed to advance the field of real-time noise suppression optimized for human perception in challenging noisy conditions. Large inclusive and diverse training and test datasets with supporting scripts were open sourced. Many participants from both industry and academia found the datasets very useful and submitted their enhanced clips for final evaluation. Only two teams participated in the personalized DNS track, which also shows that the field is in its nascent phase.
\vspace{-2mm}

\vfill\pagebreak
\bibliographystyle{IEEEbib}
\bibliography{strings,refs}

\end{document}